\documentclass[10pt,a4paper]{article}

\usepackage[utf8]{inputenc}
\usepackage[english]{babel}

\begin{document}
\large

\newpage
\begin{center}
\LARGE
{\bf A Theory of Flavor and Set of the Interaction 
\\Structural Parts}
\end{center}
\vspace{0.1cm}
\begin{center}
{\bf Rasulkhozha S. Sharafiddinov}
\end{center}
\vspace{0.1cm}
\begin{center}
{\bf Institute of Nuclear Physics, Uzbekistan Academy of Sciences,
\\Tashkent, 100214 Ulugbek, Uzbekistan}
\end{center}
\vspace{0.1cm}

\begin{center}
{\bf Abstract}
\end{center}

At the availability of a sharp interconnection, the difference in sizes
with Dirac and Pauli form factors of a massive neutrino must constitute
their linearly ordered set. Such a class of currents can lead in the field of
a spinless nucleus to the constitution of a partially ordered set consisting
of cross sections of polarized and unpolarized neutrino scattering. We discuss
a theory, in which flavor conservation is predicted as a theorem about the
equality of cross sections of the interaction with a gauge boson of leptonic
current structural components. This theorem relates flavor and mirror
symmetries as a consequence of force unification forming the two left-(right-)
handed fermions in individual difermions. Thereby, it requires one to follow
the logic of a true nature of elementary particles from the point of view of
a flavor dynamics of all types of micro-world symmetry laws.

\vspace{0.8cm}
\noindent
{\bf 1. Introduction}
\vspace{0.4cm}

Owing to the unified nature of the most diverse types of interactions with
virtual photons, the cross section of elastic scattering of neutrinos in the
nuclear Coulomb field contains all the information necessary for the creation
of a true picture of this naturally united process. Here those aspects of the
phenomenon are highly interesting which give the possibility to understand
the mathematical logic of fundamental structure of matter [1,2].

According to quantum electrodynamics, the lepton interaction with the field
of emission may be described in the form [3,4] of vector $(V_{l})$ current
\begin{equation}
j_{em}^{V_{l}}=\overline{u}(p',s')[\gamma_{\mu}f_{1l}(q^{2})-
i\sigma_{\mu\lambda}q_{\lambda}f_{2l}(q^{2})]u(p,s),
\label{1}
\end{equation}
where $\sigma_{\mu\lambda}=[\gamma_{\mu},\gamma_{\lambda}]/2,$ $p(s)$ and
$p'(s')$ imply the four-momenta (helicities) of a particle before and
after the scattering. Additionally, $q=p-p'$ is the momentum transfer.

Dirac and Pauli form factors, $f_{1l}$ and $f_{2l}$ respectively, are
responsible for the structure of the united Coulomb interactions as well
as for their unified nature. From this point of view, they constitute a
set of currents which can symbolically be written in the following way
\begin{equation}
F_{l}^{V_{l}}(q^{2})=\{f_{1l}(q^{2}), \, \, \, \, f_{2l}(q^{2})\}.
\label{2}
\end{equation}

The elements of $F_{l}^{V_{l}}$ when $q^{2}\rightarrow 0$ define the static
sizes of the electric charge and of the magnetic moment for the lepton,
respectively
\begin{equation}
e_{l}=f_{1l}(0), \, \, \, \, \mu_{l}=f_{2l}(0)
\label{3}
\end{equation}
expressed in units of electron charge $e$ and Bohr magnetons
$\mu_{B}=e/2m_{e}.$

As well as in (\ref{2}), form factors $f_{1l}$ and $f_{2l}$ in
\begin{equation}
F_{l}^{V_{l}}(0)=\{f_{1l}(0), \, \, \, \, f_{2l}(0)\}
\label{4}
\end{equation}
have different dimensionalities.

Insofar as a question about the unified theoretical description of both types
of form factors is concerned, they can appear in the interaction structure
dependence [5]. However, their explicit form has not yet been established
analytically. Therefore, it seems that $f_{2l}(0)$ describes the anomalous
magnetic moment [6], and the lepton full magnetic moment is defined [7]
analogously with Sachs [8] form factors for nucleons: 
$\mu_{l}^{full}=(f_{1l}(0)/2m_{l})+f_{2l}(0).$

In spite of a large number of works dedicated to different aspects of
the interaction between the lepton and the field of emission, there is no
unified sight of the nature of form factors in the behavior dependence of
particles of the same flavor. In the first turn, this is explained by the
absence of convincing approaches in deciding the problem from the point
of view of a kind of neutrino of each lepton.

To express the idea of any of such pairs at the fundamental dynamical level,
one must establish a true picture of the united interactions including a
unified theoretical description of all types of lepton flavors.

For our purposes, it is desirable to elucidate whether there exists any 
interratio of the contributions of $f_{1l}$ and $f_{2l}$ to the interaction 
cross section with these currents. If so, we would need to know what the 
observed connection allows us to say about the structure of sets (\ref{2}) 
and (\ref{4}) as well as the nature of the flavor symmetry 
of elementary particles.

Here we investigate this highly important question studying, in the case 
of elastic scattering on a spinless nucleus, the ideas of each interaction 
structural component in the longitudinal case of polarization of leptons and 
their massive neutrinos.

\vspace{0.8cm}
\noindent
{\bf 2. Unity of lepton vector Coulomb interaction structural parts}
\vspace{0.4cm}

The amplitude of the considered process on the basis of (\ref{1}) may
to the lower order in $\alpha$ be chosen
\begin{equation}
M^{em}_{fi}=\frac{4\pi\alpha}{q^{2}}\overline{u}(p',s')[\gamma_{\mu}
f_{1l}(q^{2})-i\sigma_{\mu\lambda}q_{\lambda}f_{2l}(q^{2})]u(p,s)
<f|J^{\gamma}_{\mu}(q)|i>,
\label{5}
\end{equation}
where $l=e,$ $\mu,$ $\tau=e_{L,R},$ $\mu_{L,R},$ $\tau_{L,R}$ or $\nu_{l}=
\nu_{e L,R},$ $\nu_{\mu L,R},$ $\nu_{\tau L,R},$ $J_{\mu}^{\gamma}$
denotes the target nucleus photon current [9].

The availability of the term $f_{1\nu_{l}}$ in (\ref{5}) would seem to
contradict gauge invariance. However, unlike the earlier presentations about
this symmetry, its mass structure [2,10] allows us to relate the mass
to charge of the neutrino as a consequence of mass-charge duality [11].

At first sight, the latter violates the charge quantization law. On the
other hand, as was noted in [12] for the fist time, any of electrically
charged particles may serve as a certain indication to the existence of a kind
of magnetically charged monoparticle. At this situation, the same mononeutrino
must lead to quantization of electric charges of all neutrinos and vice versa.

This is exactly the same as when introducing an arbitrary charge. Such a
procedure in the framework of the standard electroweak theory is, by itself,
not excluded [13].

In the presence of massive Dirac neutrinos of a vector nature, the legality
of conservation of summed charge in the decays of the neutron, muon, tau
lepton and in other processes with such neutrinos follows from the fact
that in them appear [14,15] dileptons
\begin{equation}
(l_{L}, {\bar \nu_{lR}}), \, \, \, \,
(l_{R}, {\bar \nu_{lL}}),
\label{6}
\end{equation}
\begin{equation}
(\overline{l}_{R}, \nu_{lL}), \, \, \, \,
(\overline{l}_{L}, \nu_{lR})
\label{7}
\end{equation}
and paradileptons
\begin{equation}
\{(l_{L}, {\bar \nu_{lR}}),
(\overline{l}_{R}, \nu_{lL})\}, \, \, \, \,
\{(l_{R}, {\bar \nu_{lL}}),
(\overline{l}_{L}, \nu_{lR})\}
\label{8}
\end{equation}
of a definite [16,17] flavour:
\begin{equation}
L_{l}=\left\{ {\begin{array}{l}
{+1\quad \mbox{for}\quad l_{L}, \, \, \, \, \, l_{R}, \, \, \, \,
\nu_{lL}, \, \, \, \, \nu_{lR},}\\
{-1\quad \mbox{for}\quad \overline{l}_{R}, \, \, \, \, \overline{l}_{L},
\, \, \, \, {\bar \nu_{lR}}, \, \, \, \, {\bar \nu_{lL}},}\\
{\, \, \, \, \, 0\quad \mbox{for}\quad \mbox{remaining particles.}}\\
\end{array}}\right.
\label{9}
\end{equation}

It is not surprising therefore that at the availability of the interaction
(\ref{5}), the elastic scattering cross section of leptons and their neutrinos
on an electric charge of a nucleus must have the same compound structure:
$$\frac{d\sigma_{em}^{V_l}(\theta_{l},s,s')}{d\Omega}=
\frac{1}{2}\sigma^{l}_{o}(1-\eta^{2}_{l})^{-1}\{(1+ss')f_{1l}^{2}+$$
\begin{equation}
+\eta^{2}_{l}(1-ss')[f_{1l}^{2}+4m_{l}^{2}(1-\eta^{-2}_{l})^{2}f_{2l}^{2}]
tg^{2}\frac{\theta_{l}}{2}\} F^{2}_{E}(q^{2}).
\label{10}
\end{equation}

The index $V_{l}$ implies that here there are no axial-vector currents
$A_{l},$ and
$$\sigma_{o}^{l}=
\frac{\alpha^{2}cos^{2}(\theta_{l}/2)}{4E^{2}_{l}(1-\eta^{2}_{l})
sin^{4}(\theta_{l}/2)}, \, \, \, \, \eta_{l}=\frac{m_{l}}{E_{l}},$$
$$E_{l}=\sqrt{p^{2}+m_{l}^{2}}, \, \, \, \,
F_{E}(q^{2})=ZF_{c}(q^{2}),$$
$$q^{2}=-4E_{l}^{2}(1-\eta_{l}^{2})sin^{2}\frac{\theta_{l}}{2},$$
where $F_{c}(q^{2})$ is the charge distribution form factor $(F_{c}(0)=1)$
of a nucleus with number of protons $Z,$ $\theta_{l}$ is the scattering
angle of a particle; and $E_{l}$ and $m_{l}$ are its energy and mass,
respectively.

The existence of self interference contributions $f_{il}^{2}$ of each of the
current Dirac $(i=1)$ and Pauli $(i=2)$ components for the scattering cross
section (\ref{10}) is explained by the appearance in the nuclear charge
field of the left- and right-handed [15] individual difermions:
\begin{equation}
(l_{L}, \overline{l}_{R}), \, \, \, \,
(l_{R}, \overline{l}_{L}),
\label{11}
\end{equation}
\begin{equation}
(\nu_{lL}, {\bar \nu_{lR}}), \, \, \, \,
(\nu_{lR}, {\bar \nu_{lL}}).
\label{12}
\end{equation}

As seen, left- $(s=-1)$ or right- $(s=+1)$ polarized fermion passing in
the field of a nucleus suffers either conservation $(s'=s)$ or flip $(s'=-s)$
of his helicity. We can, therefore, replace the cross section (\ref{10})
by summing size
\begin{equation}
d\sigma_{em}^{V_{l}}(\theta_{l},s)=
d\sigma_{em}^{V_{l}}(\theta_{l},f_{1l},s)+
d\sigma_{em}^{V_{l}}(\theta_{l},f_{2l},s).
\label{13}
\end{equation}

Charge and magnetic contributions have the values
$$\frac{d\sigma_{em}^{V_{l}}(\theta_{l},f_{1l},s)}{d\Omega}=
\frac{d\sigma_{em}^{V_{l}}(\theta_{l},f_{1l},s'=s)}{d\Omega}+
\frac{d\sigma_{em}^{V_{l}}(\theta_{l},f_{1l},s'=-s)}{d\Omega}=$$
\begin{equation}
=\sigma^{l}_{o}(1-\eta^{2}_{l})^{-1}(1+\eta_{l}^{2}tg^{2}
\frac{\theta_{l}}{2})f_{1l}^{2}F_{E}^{2}(q^{2}),
\label{14}
\end{equation}
$$\frac{d\sigma_{em}^{V_{l}}(\theta_{l},f_{2l},s)}{d\Omega}=
\frac{d\sigma_{em}^{V_{l}}(\theta_{l},f_{2l},s'=-s)}{d\Omega}=$$
\begin{equation}
=4m_{l}^{2}\sigma^{l}_{o}(1-\eta^{2}_{l})\eta^{-2}_{l}f_{2l}^{2}
F_{E}^{2}(q^{2})tg^{2}\frac{\theta_{l}}{2}.
\label{15}
\end{equation}

In the same way making the averaging over $s$ and summing over $s',$
we can present (\ref{10}) in the form
\begin{equation}
d\sigma_{em}^{V_{l}}(\theta_{l})=
d\sigma_{em}^{V_{l}}(\theta_{l},f_{1l})+
d\sigma_{em}^{V_{l}}(\theta_{l},f_{2l}),
\label{16}
\end{equation}
where the Dirac and Pauli cross sections are equal to
\begin{equation}
\frac{d\sigma_{em}^{V_{l}}(\theta_{l},f_{1l})}{d\Omega}=
\sigma^{l}_{o}(1-\eta^{2}_{l})^{-1}
(1+\eta^{2}_{l}tg^{2}\frac{\theta_{l}}{2})f_{1l}^{2}F_{E}^{2}(q^{2}),
\label{17}
\end{equation}
\begin{equation}
\frac{d\sigma_{em}^{V_{l}}(\theta_{l},f_{2l})}{d\Omega}=
4m_{l}^{2}\sigma^{l}_{o}(1-\eta^{2}_{l})\eta^{-2}_{l}f_{2l}^{2}
F_{E}^{2}(q^2)tg^{2}\frac{\theta_{l}}{2}.
\label{18}
\end{equation}

Thus, (\ref{13}) and (\ref{16}) would seem to say about that incoming 
neutrinos either have longitudinal polarization or are strictly unpolarized. 
This, however, is not quite so. In fact, helicity properties of a particle 
depend not only on its charge and moment [5,18], but also on the structure 
of medium [19], in which it moves. If an unpolarized neutrino interacts with
the field of emission, the latter can polarize it over its spin. Therefore, it 
should be expected that there are polarized and unpolarized particles among
the incoming and outcoming fermions. Then it is possible, for example, a flux
of the scattered neutrinos is a partially ordered set of outgoing particles.

Their scattering on nuclei is described naturally by a set of corresponding
cross sections. In the case of a nucleus with a zero spin and the neutrino
currents (\ref{1}), this class behaves as follows
$$d\sigma_{em}^{V_{l}}=
\{d\sigma_{em}^{V_{l}}(\theta_{l},f_{1l},s), \, \, \,\,
d\sigma_{em}^{V_{l}}(\theta_{l},f_{2l},s), \, \, \, \, $$
\begin{equation}
d\sigma_{em}^{V_{l}}(\theta_{l},f_{1l}), \, \, \, \,
d\sigma_{em}^{V_{l}}(\theta_{l},f_{2l})\}.
\label{19}
\end{equation}

Of course, this set becomes partially ordered [20] if between
its two elements $d\sigma_{em}^{V_{l}}(\theta_{l},f_{1l},s)$ and
$d\sigma_{em}^{V_{l}}(\theta_{l},f_{2l},s)$ another relation takes place
\begin{equation}
d\sigma_{em}^{V_{l}}(\theta_{l},f_{1l},s)\le
d\sigma_{em}^{V_{l}}(\theta_{l},f_{2l},s)
\label{20}
\end{equation}
such that
\begin{enumerate}
\item
$d\sigma_{em}^{V_{l}}(\theta_{l},f_{1l},s)\le
d\sigma_{em}^{V_{l}}(\theta_{l},f_{1l},s)$ (reflexivity),
\item
$d\sigma_{em}^{V_{l}}(\theta_{l},f_{1l},s)\le
d\sigma_{em}^{V_{l}}(\theta_{l},f_{2l},s)$ and
$d\sigma_{em}^{V_{l}}(\theta_{l},f_{2l},s)\le
d\sigma_{em}^{V_{l}}(\theta_{l},f_{1l},s)$ say
$d\sigma_{em}^{V_{l}}(\theta_{l},f_{2l},s)=
d\sigma_{em}^{V_{l}}(\theta_{l},f_{1l},s)$ (antisymmetry),
\item
$d\sigma_{em}^{V_{l}}(\theta_{l},f_{1l},s)\le
d\sigma_{em}^{V_{l}}(\theta_{l},f_{2l},s)$ and
$d\sigma_{em}^{V_{l}}(\theta_{l},f_{2l},s)\le
d\sigma_{em}^{V_{l}}(\theta_{l},f_{1l})$ imply
$d\sigma_{em}^{V_{l}}(\theta_{l},f_{1l},s)\le
d\sigma_{em}^{V_{l}}(\theta_{l},f_{1l})$ (transitivity).
\end{enumerate}

One of the most highlighted features of form factors $f_{1l}$ and $f_{2l}$ 
is their connection which involves that any dipole moment arises at the expense 
of a kind of charge [21]. As a consequence, each element of the set (\ref{19}) 
testifies in favor of the availability of all the remaining ones. Therefore,
a relation (\ref{20}) is such that 
$d\sigma_{em}^{V_{l}}(\theta_{l},f_{1l},s)\le
d\sigma_{em}^{V_{l}}(\theta_{l},f_{2l},s)$
implies $d\sigma_{em}^{V_{l}}(\theta_{l},f_{2l},s)\le
d\sigma_{em}^{V_{l}}(\theta_{l},f_{1l},s)$ (symmetry). Thus, if (\ref{20})
satisfies simultaneously the conditions both symmetry and antisymmetry, it 
must be a relation of equality.

We note also that the cross sections (\ref{13}) and (\ref{16})
are not different:
\begin{equation}
\frac{d\sigma_{em}^{V_{l}}(\theta_{l},s)}
{d\sigma_{em}^{V_{l}}(\theta_{l})}=1.
\label{21}
\end{equation}

Therefore, on the basis of (\ref{21}) one can individually compare the
contribution of any of currents $f_{1l}$ or $f_{2l}$ to the cross sections 
of longitudinal polarized and unpolarized fermion scattering. However, similar 
comparison of (\ref{13}) and (\ref{16}) establishes the identity
\begin{equation}
\frac{d\sigma_{em}^{V_{l}}(\theta_{l},f_{1l},s)}
{d\sigma_{em}^{V_{l}}(\theta_{l},f_{1l})}=1, \, \, \, \,
\frac{d\sigma_{em}^{V_{l}}(\theta_{l},f_{2l},s)}
{d\sigma_{em}^{V_{l}}(\theta_{l},f_{2l})}=1
\label{22}
\end{equation}
and thereby one can predict one more highly important regularity that class
(\ref{19}) with a partial order is of those partially ordered sets, in which
the symmetry and antisymmetry of relation (\ref{20}) lead to a relationship
implied from its transitivity. Such connections can appear in the class
(\ref{19}) even at the different permutation of elements. This reflects
the characteristic features of their structure depending on nature of the
corresponding mechanism responsible for unity of the interaction
structural parts.

These facts indicate that between the contributions of $f_{1l}$ and $f_{2l}$ 
to the cross sections both in (\ref{13}) and in (\ref{16}) there exists the 
same equality, because of which the interratios of the possible pairs of 
elements in set (\ref{19}) constitute four most diverse ratios.

To express their idea more clearly, one must apply to the two of them:
\begin{equation}
\frac{d\sigma_{em}^{V_{l}}(\theta_{l},f_{2l},s)}
{d\sigma_{em}^{V_{l}}(\theta_{l},f_{1l},s)}=1, \, \, \, \,
\frac{d\sigma_{em}^{V_{l}}(\theta_{l},f_{2l})}
{d\sigma_{em}^{V_{l}}(\theta_{l},f_{1l})}=1,
\label{23}
\end{equation}
or to the two remaining equations.

The importance of our choice lies in the fact that a non-zero Pauli interaction 
is consequence of the availability of a kind of Dirac interaction. This unites 
all elements of the class (\ref{19}) in a unified whole.

For further substantiation of their legality, we include in the discussion the 
united regularity. As in (\ref{20}), the relation
\begin{equation}
d\sigma_{em}^{V_{l}}(\theta_{l},f_{1l},s)\sim
d\sigma_{em}^{V_{l}}(\theta_{l},f_{2l},s)
\label{24}
\end{equation}
between the elements $d\sigma_{em}^{V_{l}}(\theta_{l},f_{1l},s)$ and 
$d\sigma_{em}^{V_{l}}(\theta_{l},f_{2l},s)$ in (\ref{19}) is a relation of 
equivalence [20] only if it corresponds to the conditions
\begin{enumerate}
\item $d\sigma_{em}^{V_{l}}(\theta_{l},f_{1l},s)\sim
d\sigma_{em}^{V_{l}}(\theta_{l},f_{1l},s)$ (reflexivity),
\item
$d\sigma_{em}^{V_{l}}(\theta_{l},f_{1l},s)\sim
d\sigma_{em}^{V_{l}}(\theta_{l},f_{2l},s)$ implies
$d\sigma_{em}^{V_{l}}(\theta_{l},f_{2l},s)\sim
d\sigma_{em}^{V_{l}}(\theta_{l},f_{1l},s)$ (symmetry),
\item
$d\sigma_{em}^{V_{l}}(\theta_{l},f_{1l},s)\sim
d\sigma_{em}^{V_{l}}(\theta_{l},f_{2l},s)$ and 
$d\sigma_{em}^{V_{l}}(\theta_{l},f_{2l},s)\sim
d\sigma_{em}^{V_{l}}(\theta_{l},f_{1l})$ say
$d\sigma_{em}^{V_{l}}(\theta_{l},f_{1l},s)\sim
d\sigma_{em}^{V_{l}}(\theta_{l},f_{1l})$ (transitivity).
\end{enumerate}

Similar equivalence relation separates the set (\ref{19}) to the possible
subclasses not containing the same elements. We must, therefore, replace
the class (\ref{19}) with
\begin{equation}
d\sigma_{em}^{V_{l}}=\{d\sigma_{em}^{V_{l}}(\theta_{l},s), \, \, \,\,
d\sigma_{em}^{V_{l}}(\theta_{l})\}.
\label{25}
\end{equation}

Its subsets reflect just a regularity that any of (\ref{13}) and (\ref{16})
constitutes his own set of cross sections:
\begin{equation}
d\sigma_{em}^{V_{l}}(\theta_{l},s)=
\{d\sigma_{em}^{V_{l}}(\theta_{l},f_{1l},s), \, \, \, \,
d\sigma_{em}^{V_{l}}(\theta_{l},f_{2l},s)\},
\label{26}
\end{equation}
\begin{equation}
d\sigma_{em}^{V_{l}}(\theta_{l})=
\{d\sigma_{em}^{V_{l}}(\theta_{l},f_{1l}), \, \, \, \,
d\sigma_{em}^{V_{l}}(\theta_{l},f_{2l})\}.
\label{27}
\end{equation}

Both subclasses, as follows from (\ref{21}), are strictly identical. Such a
picture may be established only in the case when the scattering cross sections
describe those processes, in which appear the same difermions of a vector or
an axial-vector nature. Each neutrino in any of them, as shown below, possesses
simultaneously only one of the C-invariant $(V_{l})$ or the C-noninvariant
$(A_{l})$ currents [1].

There exists, however, the possibility that regardless of sizes of the
interaction cross sections with the field of a nucleus of unpolarized and 
longitudinal polarized fermions, their interratio for each lepton and its 
neutrino has the same value. Under these circumstances, (\ref{13}) and 
(\ref{16}) lead to the replacement of (\ref{21}) to the following:
\begin{equation}
\frac{d\sigma_{em}^{V_{\nu_{l}}}(\theta_{\nu_{l}},s)}
{d\sigma_{em}^{V_{\nu_{l}}}(\theta_{\nu_{l}})}=
\frac{d\sigma_{em}^{V_{l}}(\theta_{l},s)}
{d\sigma_{em}^{V_{l}}(\theta_{l})}.
\label{28}
\end{equation}

This united state in turn establishes the corresponding equations
\begin{equation}
\frac{d\sigma_{em}^{V_{\nu_{l}}}(\theta_{\nu_{l}},f_{i\nu_{l}},s)}
{d\sigma_{em}^{V_{\nu_{l}}}(\theta_{\nu_{l}},f_{i\nu_{l}})}=
\frac{d\sigma_{em}^{V_{l}}(\theta_{l},f_{il},s)}
{d\sigma_{em}^{V_{l}}(\theta_{l},f_{il})}
\label{29}
\end{equation}
and thereby describes a situation when the numbers of the same elements
in the classes (\ref{26}) and (\ref{27}) for $\nu_{l}$ and $l$ coincide.

The latter convinces us here that reflexivity, symmetry and transitivity
of an equivalence relation (\ref{24}) hold independently from the type of
a particle, at which the interratio of any pair of elements from subsets
(\ref{26}) and (\ref{27}) consists of the same sizes for both $\nu_{l}$
and $l.$ They have the crucial value for establishment of the four more
most diverse relationships.

To elucidate this idea, it is desirable to use two of them:
\begin{equation}
\frac{d\sigma_{em}^{V_{\nu_{l}}}(\theta_{\nu_{l}},f_{2\nu_{l}},s)}
{d\sigma_{em}^{V_{\nu_{l}}}(\theta_{\nu_{l}},f_{1\nu_{l}},s)}=
\frac{d\sigma_{em}^{V_{l}}(\theta_{l},f_{2l},s)}
{d\sigma_{em}^{V_{l}}(\theta_{l},f_{1l},s)},
\label{30}
\end{equation}
\begin{equation}
\frac{d\sigma_{em}^{V_{\nu_{l}}}(\theta_{\nu_{l}},f_{2\nu_{l}})}
{d\sigma_{em}^{V_{\nu_{l}}}(\theta_{\nu_{l}},f_{1\nu_{l}})}=
\frac{d\sigma_{em}^{V_{l}}(\theta_{l},f_{2l})}
{d\sigma_{em}^{V_{l}}(\theta_{l},f_{1l})},
\label{31}
\end{equation}
or the two remaining equations.

Furthermore, if we consider the case $E_{l}\gg m_{l}$ when
$\eta_{l}\rightarrow 0,$ and $q^{2}\rightarrow 0$ implies
$\theta_{l}\rightarrow 0,$ the limit [15,21] arising in (\ref{23})
taking into account equations (\ref{14}), (\ref{15}), (\ref{17}) and
(\ref{18}) has the following value
$$lim_{\eta_{l}\rightarrow 0, \theta_{l}\rightarrow 0}
\frac{\eta_{l}^{2}(1+\eta_{l}^{2}tg^{2}(\theta_{l}/2))}
{(1-\eta_{l}^{2})^{2}tg^{2}(\theta_{l}/2)}=1$$
owing to this, the latter equation can be explicitly expressed as follows
\begin{equation}
2m_{l}\frac{f_{2l}(0)}{f_{1l}(0)}=\pm 1.
\label{32}
\end{equation}

Unification of (\ref{32}) with (\ref{3}) suggests a connection
\begin{equation}
f_{2l}(0)=\frac{f_{1l}(0)}{2m_{l}}=\frac{e_{l}}{2m_{l}}
\label{33}
\end{equation}
and that, consequently, $f_{2l}$ gives the normal magnetic moment of a Dirac
particle [22]. Insofar as the anomalous components of its currents are
concerned, they appear in the processes with exchange by the two bosons.

Thus, using (\ref{17}), (\ref{18}) and (\ref{22}), we can relate the masses
to a ratio of vector currents of any lepton and its neutrino, on the basis
of (\ref{30}) and (\ref{31}). This let us establish
\begin{equation}
m_{\nu_{l}}\frac{f_{2\nu_{l}}(0)}{f_{1\nu_{l}}(0)}=
\pm m_{l}\frac{f_{2l}(0)}{f_{1l}(0)}.
\label{34}
\end{equation}

Comparison of (\ref{34}) with (\ref{33}) at $l=\nu_{l}$ leads us to (\ref{32})
once more, confirming that we cannot exclude the existence of both individual
and united connections of the structural components of vector cross sections
in the lepton and its neutrino scattering.

\vspace{0.8cm}
\noindent
{\bf 3. Unification of flavor and mirror symmetries}
\vspace{0.4cm}

Returning to (\ref{30}) and (\ref{31}), we remark that each of them is
dynamically based on the definite flavor. In other words, these connections
correspond in nature to the coexistence of leptons and their neutrinos.
Consequently, the presence of any type of charged lepton implies the existence
of a neutrino of Dirac type. Such C-invariant pairs constitute the naturally
united families of the left- and right-handed leptons. Therefore, all elements
of a set such as (\ref{19}) fits answer to one of spin states of paraneutrinos,
because of which the structural components of cross sections (\ref{13}) and
(\ref{16}) coincide. Thereby, between the two particles of each of difermions
(\ref{11}) and (\ref{12}) there exists a hard flavor symmetrical
interconnection [1,15].

This principle is not changed [1] even at the interaction with virtual
photon of an axial-vector $(A_{l})$ current
\begin{equation}
j_{em}^{A_{l}}=\overline{u}(p',s')\gamma_{5}[\gamma_{\mu}
g_{1l}(q^{2})-i\sigma_{\mu\lambda}q_{\lambda}g_{2l}(q^{2})]u(p,s).
\label{35}
\end{equation}

In the framework of recent literature about the nature of Dirac fermions, 
(\ref{35}) consists of CP-symmetrical $g_{1l}$ and CP-antisymmetrical $g_{2l}$ 
parts of the different T-invariance with the same P-noninvariance. They define 
at $q^{2}\rightarrow 0$ the static anapole [23] and electric 
dipole [24] moments:
$${\it a_{l}}=\frac{1}{m_{l}}\left(\frac{g_{1l}(0)}
{f_{1l}(0)}\right)^{2}f_{2l}(0), \, \, \, \, d_{l}=g_{2l}(0).$$

The existence of $g_{1l}$ would imply also the absence of gauge invariance.
However, according to the ideas of symmetry laws, $g_{il}$ arise at the
violation of P-parity as a consequence of mass structure of gauge
invariance [2].

Finally, insofar as the weak neutral currents are concerned, they in a general
form include the two components
\begin{equation}
j_{we}=\overline{u}(p',s')\gamma_{\mu}(g_{V_{l}}+\gamma_{5}g_{A_{l}})u(p,s),
\label{36}
\end{equation}
which are characterized by the vector $g_{V_{l}}$ and axial-vector $g_{A_{l}}$
constants.

From our earlier developments [5,25], we find that the cross section
in the studied processes with partially polarized fermions in the presence
of all electroweak $(V_{l})$ and $(A_{l})$ currents (\ref{1}), (\ref{35})
and (\ref{36}) contains not only their self interference contributions
$f_{il}^{2},$ $g_{il}^{2},$ $g_{V_{l}}^{2},$ $g_{A_{l}}^{2},$
$g_{V_{l}}f_{1l}$ and $g_{A_{l}}g_{1l},$ but also the contributions
$\lambda_{c}sf_{1l}g_{1l},$ $sf_{2l}g_{2l},$ $\lambda_{c}sg_{A_{l}}f_{1l}$
and $\lambda_{c}sg_{V_{l}}g_{1l}$ of the mixed-interference [26] between
the two interactions of the vector and axial-vector nature. The value of
$\lambda_{c}$ for a particle (antiparticle) has the positive (negative) sign.
Its absence expresses, in the case of $f_{2l}g_{2l},$ the idea about that
$g_{2l}$ is C-invariant. But, as will be seen from the further, such an
implication does not corresponds to reality at all.

At first sight, such mixed-interference parts of the cross section
analogously to any self interference contribution describe the formation
of one of the discussed types of parafermions. This, however, would have
no theoretical substantiation, since the availability of the multipliers
$\lambda_{c}$ and $s$ in them is incompatible with the invariance concerning
C and P as well as with the above-mentioned equality of the scattering cross
sections of polarized and unpolarized fermions. Thus, if (\ref{21}) is valid
only for particles of the defined lepton type then flavor symmetry must be
accepted as a mirror symmetry.

\vspace{0.8cm}
\noindent
{\bf 4. On the electric charge of the C-noninvariant neutrino}
\vspace{0.4cm}

From the viewpoint of mass-charge duality [11], $f_{il}$ and $g_{il}$
constitute the vector $V_{l}$ and axial-vector $A_{l}$ parts of the same Dirac
or Pauli component of leptonic current: $j_{em}=j_{em}^{V_{l}}+j_{em}^{A_{l}}.$
In other words, the dipole moments $f_{2l}$ and $g_{2l}$ correspond to
charges [1] of a vector $f_{1l}$ and an axial-vector $g_{1l}.$

It is clear, however, that any C-even or C-odd charge may serve as the 
source of a kind of dipole [22]. Such a correspondence principle expresses 
the C-antisymmetry as well as the CP-invariance of an axial-vector electric 
dipole moment. Therefore, we do not only deny the CPT-symmetry of any type 
of C-noninvariant currents $A_{l}$ even at the violation of T-parity of the
CP-symmetrical anapole itself, but we also need to go away from the earlier
descriptions of lepton nature taking into account that the same neutrino may
not be simultaneously both a C-even fermion and a C-odd one [1,2].

In these circumstances, it seems possible to separate
all leptons into the two classes. To the first of them apply the vector
C-invariant particles. They have no axial-vector interactions. A beautiful
example is vector $V$ leptons $(l^{V}\neq \overline{l}^{V})$ and their Dirac
$(\nu_{l}^{V}\neq {\bar \nu}_{l}^{V})$ neutrinos  with $V_{l}$ currents.
The second class consists of the C-noninvariant axial-vector particles. In
them, the vector properties are absent. We include in this group the truly
neutral Majorana $(\nu_{M}^{A}={\bar \nu}_{M}^{A})$ neutrinos [27]. As the
author noticed, however, in [2] for the first time, each of these types
of neutrinos must have his own Dirac $(\nu_{D}^{A}={\bar \nu}_{D}^{A})$ 
neutrino corresponding in nature to a kind of charged lepton. A new example 
of the second group may be axial-vector $A$ leptons $(l^{A}=\overline{l}^{A})$ 
and their Dirac neutrinos with $A_{l}$ currents. Such a behavior of fermions 
leads us to the following correspondence principle
\begin{equation}
l^{V}=e^{V}, \mu^{V}, \tau^{V}, ...\rightarrow \nu_{l}^{V}=
\nu_{e}^{V}, \nu_{\mu}^{V}, \nu_{\tau}^{V}, ...,
\label{37}
\end{equation}
\begin{equation}
l^{A}=e^{A}, \mu^{A}, \tau^{A}, ...\rightarrow \nu_{l}^{A}=
\nu_{e}^{A}, \nu_{\mu}^{A}, \nu_{\tau}^{A}, ...,
\label{38}
\end{equation}
\begin{equation}
\nu_{D}^{A}=\nu_{e}^{A}, \nu_{\mu}^{A},
\nu_{\tau}^{A}, ...\rightarrow \nu_{M}^{A}=\nu_{1}^{A},
\nu_{2}^{A}, \nu_{3}^{A}, ....
\label{39}
\end{equation}

The difference in nature of truly neutral neutrinos of Dirac and Majorana types
will be treated in our further works. But here we can use, for example, any of 
earlier experiments [28] 
\newline
\noindent
about neutrino oscillations [29] as the reliable practical indication to 
the availability in all C-noninvariant particles of a kind of axial-vector 
electric charge.

\vspace{0.8cm}
\noindent
{\bf 5. Conclusion}
\vspace{0.4cm}

The presence of both $\nu_{l}^{V}({\bar \nu}^{V}_{l})$ and
$\nu_{l}^{A}({\bar \nu}^{A}_{l})$ in the field of emission leads to
a formation not only of parafermions
\begin{equation}
(\nu^{V}_{lL}, {\bar \nu}^{V}_{lR}), \, \, \, \,
(\nu^{V}_{lR}, {\bar \nu}^{V}_{lL}),
\label{40}
\end{equation}
\begin{equation}
(\nu^{A}_{lL}, {\bar \nu}^{A}_{lR}), \, \, \, \,
(\nu^{A}_{lR}, {\bar \nu}^{A}_{lL})
\label{41}
\end{equation}
of the same vector $V$ or an axial-vector $A$ nature, but also of the united
systems of the two Dirac neutrinos of the different C-invariance
\begin{equation}
(\nu^{V}_{lL}, {\bar \nu}^{A}_{lR}), \, \, \, \,
(\nu^{V}_{lR}, {\bar \nu}^{A}_{lL}),
\label{42}
\end{equation}
\begin{equation}
(\nu^{A}_{lL}, {\bar \nu}^{V}_{lR}), \, \, \, \,
(\nu^{A}_{lR}, {\bar \nu}^{V}_{lL}).
\label{43}
\end{equation}

The interaction with a nucleus of each paraneutrino from (\ref{42}) and
(\ref{43}) violates, as has been mentioned above, the equality (\ref{21})
expressing the idea of lepton number conservation law. Such a violation can
explain once again the absence of earlier known lepton flavors for all types
of truly neutral leptons and their neutrinos.

These reasonings give the right to interpret the flavor symmetry as 
a unification of the two left- (right-) handed fermions in individual 
difermions of $V_{l}$ or $A_{l}$ currents.

So, we have learned that (\ref{22}) and (\ref{23}) similarly to any of
(\ref{29})-(\ref{31}) correspond in nature to the same flavor. In other
words, they must be valid only for particles of the defined types of lepton
numbers, at which a class (\ref{19}) becomes partially ordered. Thus, flavor
conservation in the elastic scattering must be considered as a theorem about
the equality of cross sections of the interaction with the field of emission
of leptonic current structural parts.

According to one of the dynamical aspects of this theorem, each of equations
in (\ref{23}) does not differ from unit and thereby it is assumed that either
$f_{1l}<f_{2l}$ or $f_{2l}<f_{1l}.$

At these situations, sets (\ref{2}) and (\ref{4}) become linearly ordered. 
But their all the elements may not be defined completely. There exists a range 
of uncertainties both in the behavior and in the structure of leptonic current.
Another reason of incompleteness is the absence of a true picture of phenomenon
from the point of view of spin polarization. Nevertheless, taking into account
the above-noted regularities of fermion nature, we conclude that a linearly
ordered class of massive neutrino currents in the field of emission
constitutes a partially ordered set of cross sections.

Thus, unlike the earlier principles of the united interactions, our theorem
about unification of forces requires the establishment of nature of elementary
particles in the flavor structure dependence of fundamental symmetry laws.

\vspace{0.8cm}
\noindent
{\bf References}
\begin{enumerate}
\item
R.S. Sharafiddinov, Fizika {\bf B 16} (2007) 1;
hep-ph/0512346.
\item
R.S. Sharafiddinov, Phys. Essays {\bf 19} (2006) 58;
hep-ph/0407262.
\item
G. Feinberg and L.M. Lederman, Ann. Rev. Nucl.
Sci. {\bf 13} (1963) 157.
\item
R.S. Sharafiddinov, Spacetime Subst. {\bf 5} (2004) 32;
hep-ph/0306145.
\item
R.S. Sharafiddinov, Spacetime Subst. {\bf 3} (2002) 134;
physics/0305015.
\item
J. Schwinger, Phys. Rev. {\bf 76} (1949) 790.
\item
B.K. Kerimov, T.R. Aruri and M.Ya. Safin,
Izv. Acad. Nauk SSSR. Ser. Fiz. {\bf 37} (1973) 1768.
\item
R.G. Sachs, Phys. Rev. {\bf B 136} (1962) 281.
\item
T.W. Donnelly and R.D. Peccei, Phys. Rep. {\bf 50} (1979) 3.
\item
R.S. Sharafiddinov, Spacetime Subst. {\bf 5} (2004) 83;
hep-ph/0306255.
\item
R.S. Sharafiddinov, Spacetime Subst. {\bf 3} (2002) 47;
physics/0305008.
\item
R.S. Sharafiddinov, Spacetime Subst. {\bf 3} (2002) 132;
physics/0305014.
\item
I.S. Batkin and M.K. Sundaresan, J. Phys. {\bf G 20} (1994) 1749.
\item
R.S. Sharafiddinov, in {\it Proc. April Meeting of the American
Physical Society, Dallas, Texas, April 22-25, 2006,} Abstract, H12.00009.
\item
R.S. Sharafiddinov, hep-ph/0511065.
\item
Ya.B. Zel'dovich, Dokl. Akad. Nauk SSSR. {\bf 86} (1952) 505.
\item
E.J. Konopinsky and H. Mahmoud, Phys. Rev. {\bf 92} (1953) 1045.
\item
R.S. Sharafiddinov, Dokl. Akad. Nauk Ruz. Ser. Math.
Tehn. Estest. {\bf 7} (1998) 25; hep-ph/0307083.
\item
R.S. Sharafiddinov, hep-ph/0409254.
\item
R.R. Stoll, {\it Set Theory and Logic}, Mineola, New York, Dover
Publications (1979).
\item
R.S. Sharafiddinov, Spacetime Subst. {\bf 1} (2000) 176;
hep-ph/0305009.
\item
R.S. Sharafiddinov, Spacetime Subst. {\bf 3} (2002) 86;
physics/0305009.
\item
Ya.B. Zel'dovich, JETP {\bf 33} (1957) 1531; Ya.B. Zel'dovich and 
A.M. Perelomov, JETP {\bf 39} (1960) 1115.
\item
L.D. Landau, JETP {\bf 32} (1957) 405; Nucl. Phys. {\bf 3} (1957) 127.
\item
B.S. Yuldashev and R.S. Sharafiddinov, Spacetime Subst.
{\bf 5} (2004) 137; hep-ph/0510080.
\item
R.B. Begzhanov and R.S. Sharafiddinov, Mod. Phys.
Lett. A 15 (2000) 557.
\item
E. Majorana, Nuovo Cimento {\bf 14} (1937) 171.
\item
Y. Fukuda et. al., Phys. Rev. Lett. {\bf 81} (1998) 1562.
\item
B. Pontecorvo, JETP {\bf 33} (1957) 549.
\end{enumerate}

\end{document}